\documentclass[prd,amsfonts,aps,nofootinbib,notitlepage,10pt]{revtex4-1}

\pdfoutput=1
\usepackage{amsmath,amssymb}
\usepackage[utf8]{inputenc}
\usepackage{graphicx}
\usepackage[colorlinks=true]{hyperref}

\begin{document}
\title{The Inert Zee Model}
\author{Robinson Longas}%
\email{robinson.longas@udea.edu.co}
\author{Dilia Portillo}%
\email{dilia.portillo@udea.edu.co}
\author{Diego Restrepo}%
\email{restrepo@udea.edu.co}
\author{Oscar Zapata}%
\email{oalberto.zapata@udea.edu.co}
\affiliation{
Instituto de F\'isica, Universidad de Antioquia, Calle 70 No. 52-21, Medell\'in, Colombia}
\date{\today}

\begin{abstract}
We study a realization of the topology of the Zee model for the generation of neutrino masses at one-loop with a minimal set of vector-like fermions.  
After imposing an exact $Z_2$ symmetry to avoid tree-level Higgs-mediated flavor changing neutral currents, one dark matter candidate is obtained from the subjacent inert doublet model, 
but with the presence of new co-annihilating particles. 
We show that the model is consistent with the constraints coming from lepton flavor violation processes, oblique parameters, dark matter and neutrino oscillation data.  
\end{abstract}

 \maketitle
\section{Introduction}
\label{sec:introduction}

Neutrino masses and dark matter (DM) represent two phenomenological pieces of evidence for physics beyond the Standard Model  (SM) which are solidly supported by the experimental data. 
If neutrino masses arise radiatively~\cite{Ma:1998dn,Bonnet:2012kz,Sierra:2014rxa} it may be, though,  that both originate from new physics at the TeV scale, and they are related to each other.
In this direction, models with one-loop radiative neutrino masses and viable dark matter candidates have now a complete
classification given in~\cite{Law:2013saa,Restrepo:2013aga}. There, the  new fields are odd under a $Z_2$ symmetry which ensures the stability of the DM particle, while the SM particles
are even.  In this work, we explore a particular model where the $Z_2$ can be identified with the symmetry used to avoid tree-level Higgs-mediated flavor changing neutral currents (HMFCNC) in the 
two Higgs doublet models (THDM)~\cite{Branco:2011iw}.  More concretely, we consider the realization of the $d=5$ Weinberg operator at one-loop order~\cite{Ma:1998dn,Bonnet:2012kz} with the topology 
labeled as T1-ii  in~\cite{Bonnet:2012kz} from which the Zee model~\cite{Zee:1980ai} is the most straightforward realization.  In the Zee model, the THDM-III with tree-level HMFCNC  is extended with one extra  
$\operatorname{SU}(2)$-singlet charged-scalar. The minimal realization with two Higgs doublets of opposite parity under a $Z_2$ symmetry to avoid tree-level  HMFCNC~\cite{Wolfenstein:1980sy}, 
gives rise to a diagonal-zero neutrino mass texture which is excluded from the measurement of a non-maximal solar neutrino mixing~\cite{He:2003ih}. 

In this work, we extend the Zee model with a minimal set of vector-like (VL) fermions, consisting in a $\operatorname{SU}(2)_L$-singlet and a doublet. Then, 
we show that a consistent model without tree-level HMFCNC can be obtained if  we impose a $Z_2$ symmetry to generate the Inert Doublet Model (IDM)~\cite{Deshpande:1977rw}. 
Due to the mixing of the  two resulting charged fermions, the neutral fermion cannot be the lightest $Z_2$-odd particle, and therefore, the DM candidate is still contained in the IDM sector of the model. 
In our setup, the imposed $Z_2$ guarantees the  absence of strongly constrained flavor violating processes, relating one-loop neutrino masses with dark matter through new physics at TeV scale that 
can be tested at the LHC.

Another example of this kind of relation arises in the well known scotogenic models. There, the SM is increased with at least two singlet~\cite{Ma:2006km} or triplet~\cite{Ma:2008cu,Chao:2012sz} 
fermions and one scalar doublet which are odd under a $Z_2$ symmetry. In another realization, the roles are interchanged with at least two scalar singlets and one VL doublet fermion, while one additional 
fermion singlet is required to close the neutrino mass loop~\cite{Fraser:2014yha,Restrepo:2015ura}. The role of the $Z_2$ in the scotogenic models is to  forbid tree-level contributions to the neutrino masses
which are generated at one-loop level. In these models, the lightest odd particle (either scalar or fermion) can be a good DM candidate.
One shared feature with the model presented here is that all the new states beyond the SM are odd under the imposed $Z_2$. 
Under this assumption, and considering new fermion and scalar fields transforming as singlets, doublets or triplets of $\operatorname{SU}(2)$, a set of 35 non-equivalent models that can simultaneously account 
for DM and neutrino masses at one-loop was obtained in Ref.~\cite{Restrepo:2013aga}\footnote{A comprehensive list of the radiative seesaw literature is given in~\cite{Okada:2015nca}.}. 
The model presented here is cataloged there as the T1-ii-A model with $\alpha=-2$.

This paper is organized as follows. In section~\ref{sec:model}, we present the model with its particle content and calculate the neutrino masses. Then, we analyze the DM phenomenology and establish the 
requirements over the free parameters necessary to reproduce the IDM phenomenology. 
In section~\ref{sec:constraints}, we study the constraints coming from oblique parameters and present the expression for the rate of the $\mu \rightarrow e \gamma$ process.
In section~\ref{sec:numer-results-disc}, we present the numerical results and discuss the collider limits on VL fermions. Finally, the conclusions are presented in section~\ref{sec:conclusions}. 
In the appendices, we collect the loop functions for the calculation of the oblique parameters and the $\mu \rightarrow e \gamma$ process.

\section{The Model}
\label{sec:model}
We start as in \cite{Zee:1980ai} by extending the SM with a second Higgs doublet, $H_2$, and a charged $\operatorname{SU}(2)$-singlet, $S^+$. Within this setup,  Majorana neutrino masses are generated at one-loop. 
In this way, the Zee model is realized in the context of the general THDM-III with tree-level HMFCNC.  In the model, ten new couplings are directly related to the neutrino sector. 
In particular, the analysis in terms of THDM-III basis independent parameters~\cite{Davidson:2005cw} was done in~\cite{AristizabalSierra:2006ri}, with further analysis in~\cite{He:2011hs,Fukuyama:2010ff}.

To avoid HMFCNC at tree-level, in the Zee-Wolfenstein model~\cite{Wolfenstein:1980sy} was proposed the usual $Z_2$ symmetry in which the two doublets have opposite parity,  
like in Type-I or Type-II or other THDM realizations~\cite{Branco:2011iw}.
Under this symmetry, the Lagrangian relevant for the neutrino mass generation
requires $S^\pm$ to be $Z_2$-even,  and hence a  $Z_2$ soft-breaking mass parameter needs to be introduced in the scalar sector,
which, in joint with the three antisymmetric Yukawa 
couplings of $S^\pm$  with the lepton doublets of different families,  account for only four new couplings directly related to the neutrino sector. 
This minimal model, however, turns to be not enough to fit the observables related to neutrino oscillation data and is now excluded~\cite{He:2003ih}.  

In this work, we want to explore the minimal realization of the T1-ii topology of~\cite{Bonnet:2012kz}, which is safe regarding strongly constrained lepton-flavor violation, in particular,  
without tree-level HMFCNC. We start by assigning  a $Z_2$-odd charge to both $S^\pm$ and the second Higgs doublet $H_2$. At this level, the resulting model would be a  Type-I THDM with an extra $S^\pm$  
and massless neutrinos.  After that,  we propose one minimal extension of this setup  that only involves six additional Yukawa-couplings related to neutrino physics (instead of the nine of the general 
Zee model without the $Z_2$). This consists of adding a  $Z_2$-odd  pair of  VL fermions: a $SU(2)_L$-singlet, $\epsilon$, and  a doublet, $\Psi$.
However, the $Z_2$ symmetry is not enough to avoid mixing of the new VL fermions with the SM leptons which could regenerate tree-level HMFCNC, as well as other lepton flavor violating processes subject to 
several (stringent) constraints~\cite{Blankenburg:2012ex,Harnik:2012pb,Dermisek:2015oja,Botella:2015hoa,Altmannshofer:2015qra}. Therefore, we impose in addition that the neutral part of $H_2$ does not 
develop a vacuum expectation value (vev). In this way, the IDM is obtained, which includes a potential scalar DM candidate. To our knowledge, the model was first proposed in the catalog of the realization 
of the $d=5$ Weinberg operator at one-loop  with DM candidates~\cite{Restrepo:2013aga} and labeled there as T1-ii-A model with $\alpha=-2$. 

The new particle content and their charges are summarized  in the Table \ref{tab:content}. A similar approach with controlled FCNC and DM was followed  in~\cite{Kanemura:2015maa} where the minimal 
supersymmetric standard model was extended with two  $\operatorname{SU}(2)$-singlet opposite-charge superfields.
 
\begin{table}[h!]
\begin{center}
\begin{tabular}{|c|c|c|}
\hline
\hline\rule[0cm]{0cm}{.9em}
   & \text{Spin} & $SU(3)_C,\,SU(2)_L,\,U(1)_Y,\,Z_2$ \\ 
\hline
\hline
$\epsilon$ & 1/2 & $({\bf 1},{\bf 1},-2,-)$ \\\hline
$\Psi$ & 1/2 & $({\bf 1},{\bf 2},-1,-)$ \\\hline
$H_2$ & 0 & $({\bf 1},{\bf 2},1,-)$ \\\hline
$S^{-}$ & 0 & $({\bf 1},{\bf 1},-2,-)$ \\\hline
\end{tabular}
\end{center}
\caption{The new particle content of the model with their transformation properties under the $SU(3)_C\otimes SU(2)_L\otimes U(1)_Y {\otimes} Z_2$ symmetry.}
\label{tab:content}
\end{table}

\subsection{The scalar sector}
The most general $Z_2$-invariant scalar potential of the model is given by
\begin{align}
  \label{eq:potential}
  V & =  \mu_1^2 H_1^{\dagger}H_1 + \mu_2^2 H_2^{\dagger}H_2 + \frac{\lambda_1}{2} ( H_1^{\dagger}H_1 )^2 
  + \frac{\lambda_2}{2}  (H_2^{\dagger}H_2 )^2 
  +  \lambda_3 ( H_1^{\dagger}H_1 )( H_2^{\dagger}H_2 )
  +  \lambda_4 ( H_1^{\dagger}H_2 )( H_2^{\dagger}H_1 ) 
  \\
  & +  \frac{\lambda_5}{2}\left[( H_1^{\dagger} H_2 )^2 + {\rm h.c.} \right]  + \mu_S^2 S^+ S^- 
  + \lambda_S (S^+ S^-)^2 + \lambda_6 (S^+ S^-) H_1^{\dagger}H_1
  + \lambda_7 (S^+ S^-)(H_2^{\dagger}H_2)   
  +  \mu  \epsilon_{ab}\left[H_1^a H_2^b S^- + {\rm h.c.} \right], \nonumber
\end{align}
where $\epsilon_{ab}$ is the $SU(2)_L$ antisymmetric tensor with $\epsilon_{12}=1$, 
$H_1=(0, H_1^0)^{\text{T}}$ is the SM Higgs doublet and $H_2=(H_2^+, H_2^0)^{\text{T}}$. 
The scalar couplings $\lambda_5$ and $\mu$ are taken to be real. 
After the electroweak symmetry breaking, the neutral scalar fields can be parametrized in the form $H_2^0=(H^0 + i A^0)/\sqrt{2}$ and $H_1^0=(h+v)/\sqrt{2}$, 
with $h$ being the Higgs boson and $v=246$ GeV. Note that $H_2^0$ does not develop a vacuum expectation value in order to ensure the conservation of the $Z_2$ symmetry. 
The neutral scalar spectrum coincides with the one of the IDM~\cite{Deshpande:1977rw,Barbieri:2006dq,LopezHonorez:2006gr}, which consists of two CP-even neutral states $(H^0,\, h)$ and a CP-odd neutral state $(A^0)$. 
The masses of the $Z_2$-odd neutral scalar particles read as
\begin{align} \label{eq:scalar_neutralsmasss}
m_{H^{0}}^{2}=\mu_{2}^{2}+\frac{1}{2}\left(\lambda_{3}+\lambda_{4}+\lambda_{5}\right)v^2,\hspace{1cm} m_{A^{0}}^{2}&=\mu_{2}^{2}+\frac{1}{2}\left(\lambda_{3}+\lambda_{4}-\lambda_{5}\right)v^2.
\end{align}
On the other hand, the charged scalar sector involves a mixture of the singlet and doublet $Z_2$-odd charged states which leads to the following mass matrix in the basis $(H_2^{\pm}, S^{\pm})$  
\begin{align}
  \label{eq:scalarchargedmix}
  \mathcal{M}^2_{S} = \begin{pmatrix}
    m_{H^{\pm}}^2 & \frac{-\mu v}{\sqrt{2}}  \cr
    \frac{-\mu v}{\sqrt{2}} & m_{S^{\pm}}^2 
  \end{pmatrix},
\end{align}
where $m_{H^{\pm}}^2 = \mu_2^2 + \frac{1}{2}\lambda_3 v^2$ and $m_{S^{\pm}}^2 = \mu_S^2 + \frac{1}{2}\lambda_6 v^2$. 
The mass eigenstates $\kappa_1^\pm$ and $\kappa_2^\pm$ are defined as 
\begin{align}
  \label{eq:physical_scalars}
  \left(\begin{array}{c}
      H_2^{\pm} \\
      S^{\pm}
    \end{array}\right) = \begin{pmatrix}
    \cos\delta & -\sin\delta \cr
    \sin\delta & \cos\delta
  \end{pmatrix}
  \left(\begin{array}{c}
      \kappa_1^\pm \\
      \kappa_2^\pm
    \end{array}\right),
\hspace{1cm}\sin{2\delta}=\frac{\sqrt{2} \mu v}{m_{\kappa_2^+}^2-m_{\kappa_1^+}^2},
\end{align}
with the corresponding masses
\begin{align}
  \label{eq:schmasses}
  m_{\kappa_{1,2}^{\pm}}^2 & = 
  \frac{1}{2} \left( m_{H^{\pm}}^2 + m_{S^\pm}^2 \mp 
    \sqrt{ \left( m_{H^{\pm}}^2 - m_{S^\pm}^2 \right)^2 + 2\mu^2v^2} \right),
\end{align}
with $\mu$ constrained from above by the requirement of having $m_{\kappa_1^+}^2>0$. 

Lastly, the scalar couplings are subject to perturbativity and vacuum stability constraints, 
which imply the following conditions~\cite{Deshpande:1977rw, Kanemura:2000bq}:
\begin{align}
  \label{eq:contraints_scalarparameters}
  &\mu_1^2<0,\;\; \lambda_1\mu_2^2>\lambda_3\mu_1^2,\;\; \lambda_1\mu_S^2>\lambda_6\mu_1^2,\;\;
  \lambda_1\mu_2^2>\left(\lambda_3+\lambda_4\pm|\lambda_5|\right)\mu_1^2,\;\;|\lambda_S|,\,|\lambda_i| < 8\pi \;, \nonumber \\
  & \lambda_1,\,\lambda_2,\,\lambda_S > 0\;,\;\;   \lambda_6 > - \sqrt{\frac{\lambda_1 \lambda_S}{2}}\;,\;\; 
   \lambda_7 > - \sqrt{\frac{\lambda_2 \lambda_S}{2}} \;,\;\;
  \lambda_3 + \lambda_4 - | \lambda_5 | + \sqrt{\lambda_1 \lambda_2}> 0\;.
\end{align}
These theoretical conditions constrain the mass splittings among the $Z_2$-odd scalar particles. 
 
With regard to the free parameters in the scalar sector, it is possible to choose the following set
\begin{align}
  m_{H^0}, m_{A^0}, m_{\kappa_1^+}, m_{\kappa_2^+}, \lambda_L, \lambda_6, \delta,
\end{align}
where $\lambda_L=\frac{1}{2}\left(\lambda_{3}+\lambda_{4}+\lambda_{5}\right)$ controls the trilinear coupling between the SM Higgs and $H^0$. Because the quartic couplings $\lambda_2$, $\lambda_S$ and $\lambda_7$ 
are only relevant for interactions exclusively involving  $Z_2$-odd particles, they can be left apart in a tree-level analysis\footnote{Note that at one-loop level $\lambda_2$ and $\lambda_7$ may play a main role in 
processes such as the DM annihilations into $\gamma\gamma$ and $Z\gamma$ \cite{Gustafsson:2007pc,Garcia-Cely:2015khw}, DM scattering on nucleons \cite{Abe:2015rja} 
and other radiative processes \cite{Goudelis:2013uca}.}. The relation between the remaining scalar couplings and the scalar masses are presented in the Appendix \ref{sec:freeparameters}.  
From Eqs. (\ref{eq:scalar_neutralsmasss}) and (\ref{eq:schmasses}), we can expect that for appropriate scalar couplings, $H^0$ or $A^0$ can be the lightest $Z_2$-odd scalar particle in the scalar spectrum.

\subsection{ Yukawa interactions and the $Z_2$-odd fermion sector}
The $Z_2$-invariant Lagrangian respecting the SM gauge symmetry contains the following new terms
\begin{align}
  \label{eq:YukawaLagrangian}
  -\mathcal{L}\supset  \left\{\eta_i\bar{L}_{i}H_2\epsilon
    + \rho_i \bar{\Psi}H_2 e_{Ri} + \Pi \bar{\Psi} H_1 \epsilon 
    + f^*_i \overline{L^c_{i}} \Psi S^+  + {\rm h.c}\right\}
    + m_{\Psi} \bar{\Psi}\Psi+ m_{\epsilon} \bar{\epsilon} \epsilon \;,
\end{align}
where  $L_i$ and $e_{Ri}$ are the lepton doublets and $SU(2)$-singlets respectively, $\Psi=(N, E)^{\text{T}}$ is the VL doublet, 
$\Pi$, $\eta_i$, $\rho_i$ and $f_i$ are Yukawa-couplings controlling the new interactions, and $i$ is the family index.
As it will be shown below, the $\eta_i$, $f_i$ terms with the mixing terms $\Pi$ and $\mu$ give rise to nonzero 
neutrino masses at one loop level, and along with the $\rho_i$ term, induce lepton flavor violation (LFV)
processes such as $\mu\to e \gamma$.      

Once the electroweak symmetry is spontaneously broken the $\Pi$ term generates a mixture of the two charged $Z_2$-odd fermions, leading to a mass matrix in the basis $(E, \epsilon)$ 
given by\footnote{For simplicity we have assumed $\Pi$ to be real.}
\begin{align}
  \label{eq:chargedmix}
  \mathcal{M} = \begin{pmatrix}
    m_{\Psi} & \frac{\Pi v}{\sqrt{2}}  \cr
    \frac{\Pi v}{\sqrt{2}} & m_{\epsilon}
  \end{pmatrix} \;.
\end{align}
The charged mass eigenstates $\chi_1$ and $\chi_2$ are defined by 
\begin{align}
  \label{eq:physical_chargedfermions}
  \left(\begin{array}{c}
      E \\
      \epsilon
    \end{array}\right) = \begin{pmatrix}
    \cos\alpha & -\sin\alpha \cr
    \sin\alpha & \cos\alpha
  \end{pmatrix}
  \left(\begin{array}{c}
      \chi_1 \\
      \chi_2
    \end{array}\right)\;,\hspace{1cm}  \sin{2\alpha} & = \frac{\sqrt{2} \Pi v}{m_{\chi_2} - m_{\chi_1}},
\end{align}
with masses 
\begin{align}
  \label{eq:chargedanglesandmasses}
  m_{\chi_{1,2}} & = 
  \frac{1}{2} \left( m_{\Psi} + m_{\epsilon} \mp \sqrt{ \left( m_{\Psi} - m_{\epsilon} \right)^2 + 2\Pi^2v^2} \right).
\end{align}
The $Z_2$-odd fermion spectrum also contains a neutral Dirac fermion $N$, with a mass $m_N = m_{\Psi}$. From above expression, it follows that $m_N = m_{\chi_1}\cos^2\alpha + m_{\chi_2}\sin^2\alpha$, 
which implies the hierarchical spectrum $m_{\chi_1}\leq m_N\leq m_{\chi_2}$. In other words, the neutral fermion $N$ can not be the lightest $Z_2$-odd particle in the spectrum.

\subsection{Neutrino masses}
The usual lepton number ($L$) assignment in the Zee model corresponds to $L(H_2)=0$ and $L(S)=-2$, which makes the $\mu$ term in the scalar potential the only explicit $L$-violating term in the Lagrangian. 
Hence, by keeping such assignment and charging under $L$ the new fermion fields as $L(\Psi)=L(\epsilon)=+1$, in order to make the Yukawa interactions $L$ conserving, the $\mu$ term is again the responsible for the $L$ breaking in the model, and the subsequent neutrino Majorana masses and lepton flavor violation processes.

\begin{figure}[h!]
\begin{center}
\includegraphics[scale=0.6]{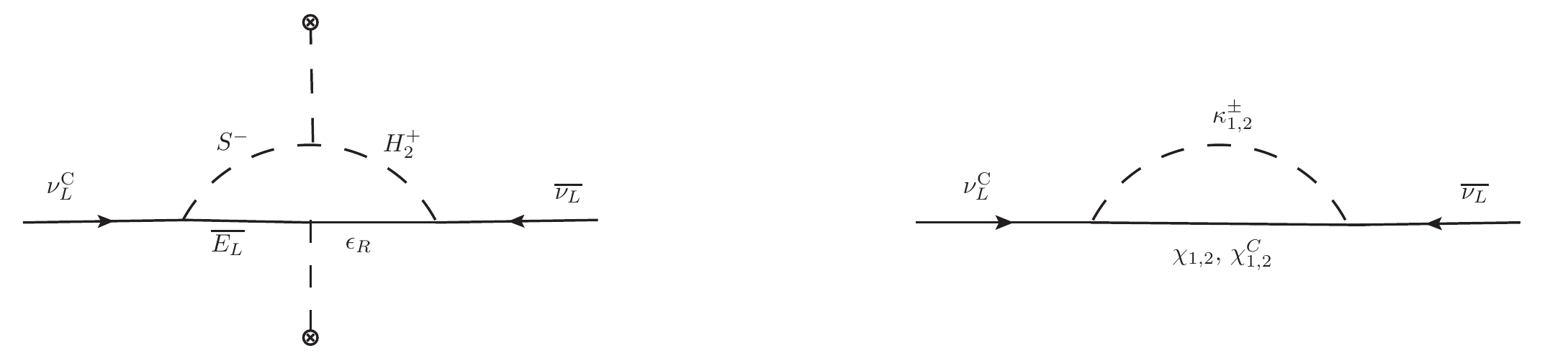}
\caption{One-loop diagram for neutrino masses in the interaction (left-panel) and mass (right-panel) basis. }
\label{fig:neutrinomass}
\end{center}
\end{figure}

Non-zero neutrino masses at one-loop are generated in this model thanks to the combination of the Yukawa-coupling  $\eta_i$ and $f_i$,  the scalar mixing $\mu$, and fermion mixing  $\Pi$,  as displayed in the 
left-panel of Fig.~\ref{fig:neutrinomass}. The corresponding Majorana mass-matrix in the mass-eigenstate basis, calculated from the Feynman diagram displayed in the right-panel of Fig.~\ref{fig:neutrinomass}, 
takes the form    
\begin{align}
 \label{eq:neutrinomass}
 [M^{\nu}]_{ij} &= \frac{\sin2\alpha \sin2\delta}{64 \pi^2} \left(\eta_i f_j + \eta_j f_i \right)\sum_nc_nm_{\chi_n}I\left(m_{\kappa_1^+}^2,m_{\kappa_2^+}^2,m_{\chi_n}^2\right).
\end{align}
Here $c_1=-1$, $c_2=+1$ and the loop function is given by 
\begin{align}
 \label{eq:loopfunction}
 I\left(m_a^2,m_b^2,m_c^2\right) &= \frac{m_b^2}{m_b^2-m_c^2}\ln\left(\frac{m_b^2}{m_c^2}\right) - \frac{m_a^2}{m_a^2-m_c^2}\ln\left(\frac{m_a^2}{m_c^2}\right)\;.
\end{align} 
Due to the flavor structure of $M^{\nu}$, it has a zero determinant and, therefore, contains only two massive neutrinos. In this way, the number of Majorana phases is reduced to only one, and neutrinos masses 
are entirely set by the solar and atmospheric mass differences. Specifically, for normal hierarchy (NH) $m_1=0$, $m_2=\sqrt{\Delta m_{\text{sol}}^2}$ and $m_3=\sqrt{\Delta m_{\text{atm}}^2}$ while for inverted 
hierarchy (IH) $m_1=\sqrt{\Delta m_{\text{atm}}^2}$, $m_2=\sqrt{\Delta m_{\text{sol}}^2+m_1^2}\approx\sqrt{\Delta m_{\text{atm}}^2}$ and $m_3=0$. On the other hand, $M^\nu$ depends on the scalar and fermion mixing 
angles with vanisihing entries for either $m_{\kappa_1^+}=m_{\kappa_2^+}$, or $m_{\chi_1}=m_{\chi_2}$. Thus, to have small neutrino masses a degenerate mass spectrum up to some extent could be required. 
By taking the trace of $M^{\nu}$ we can estimate the values of the different quantities involved in the calculation of neutrino masses:
\begin{align}
\mbox{Tr}[M^{\nu}]\approx\sqrt{\Delta m^2_{\text{atm}}}&=0.03\,\mbox{eV}\,\left(\frac{\sin2\alpha \sin2\delta}{10^{-2}}\right)\left(\frac{|\vec{\eta} \cdot \vec{f}|}{10^{-6}}\right)\sum_nc_n\left(\frac{m_{\chi_n}}{100\,\mbox{GeV}}\right)I\left(m_{\kappa_1^+}^2,m_{\kappa_2^+}^2,m_{\chi_n}^2\right).
\end{align}
This means that barring cancellations in the mass sector, and between Yukawa-couplings, small mixing angles and Yukawa-couplings are required.  Certainly large values for the Yukawa-couplings can be obtained 
for smaller values of $\sin2\alpha \sin2\delta$ or more compressed mass spectra.
  
The neutrino mass matrix is diagonalized by the Pontecorvo-Maki-Nakagawa-Sakata mixing matrix $U_{\text{PMNS}}$ \cite{Maki:1962mu} as 
\begin{align}\label{eq:neutinosdiagonal}
U_{\text{PMNS}}^{\text{T}}M^{\nu}U_{\text{PMNS}} ={\rm diag}(m_1,m_2,m_3),\hspace{1cm} m_i\ge0,
\end{align}
which can be written in the form $U_{\text{PMNS}}=VP$ \cite{Agashe:2014kda}, where the matrix $V$ contains the neutrino mixing angles and the CP Dirac phase and $P=\mbox{diag}(1,e^{i\alpha/2},1)$ 
carries the dependence on the CP Majorana phase. 
It is worth mentioning that for $\alpha=0,\pm\pi,\pm2\pi$, the Majorana phase does not contribute to the CP violation and in such a case the relative CP-parity of the two massive neutrinos
would be $\lambda=e^{\pm i\alpha}=\pm1$.  From Eq. \eqref{eq:neutinosdiagonal} and thanks to the flavor structure of the neutrino mass matrix, given by Eq. \eqref{eq:neutrinomass}, 
we can express five of the six Yukawa-couplings $\eta_i$ and $f_i$ in terms of the neutrino observables. 
Without loss of generality $\eta_1$ can be chosen to be the free parameter which can be restricted using other low energy observables such as $\mu \rightarrow e \gamma$. 
Thus, the most general Yuwawa-couplings that are compatible with the neutrino oscillation data are given by  
\begin{align}
\label{eq:yukacouplings}
 \vec{\eta} = \left|\eta_1\right| \begin{pmatrix}
                e^{i\phi_{\eta_1}} \cr
                 A_2/\beta_{11}\cr
                 A_3/\beta_{11}
                \end{pmatrix},\;\;
 \vec{f} =  \frac{1}{2\zeta} \begin{pmatrix}
                 \beta_{11}/\eta_1 \cr
                 \beta_{22}/\eta_2\cr
	         \beta_{33}/\eta_3
                \end{pmatrix}=
                \frac{\beta_{11}}{2\zeta\left|\eta_1\right|}  \begin{pmatrix}
                 e^{i\phi_{\eta_1}} \cr
                 \beta_{22}/A_2 \cr
	         \beta_{33}/A_3
                \end{pmatrix},         
\end{align}
where we have defined  
\begin{align}\label{eq:yukNH}
\beta_{ij} = \lambda m_2V_{i2}^*V_{j2}^*+ m_3V_{i3}^*V_{j3}^*,\,\,\,& A_{j} = \pm\sqrt{-\lambda m_2m_3(V_{12}^*V_{j3}^*-V_{13}^*V_{j2}^*)^2}+\beta_{1j}e^{i\text{Arg}(\eta_1)},\,\,\,  \mbox{for NH}, \\
\beta_{ij} = m_1V_{i1}^*V_{j1}^*+ \lambda m_2V_{i2}^*V_{j2}^*,\,\,\,& A_{j} = \pm\sqrt{-\lambda m_1m_2(V_{11}^*V_{j2}^*-V_{12}^*V_{j1}^*)^2}+\beta_{1j}e^{i\text{Arg}(\eta_1)},\,\,\, \mbox{for IH}, \\
&\zeta = \frac{\sin2\alpha \sin2\delta}{64 \pi^2}\sum_nc_nm_{\chi_n}I\left(m_{\kappa_1^+}^2,m_{\kappa_2^+}^2,m_{\chi_n}^2\right).
\end{align}
In this way, it is always possible to correctly reproduce the neutrino oscillation parameters in the present model. Note that, in general, the non-free Yukawa-couplings are complex numbers. 
However, they become real in a CP-conserving scenario with $\lambda=-1$ and $\eta_1$ being real. 
\subsection{Dark Matter}
\label{sec:DM}
The $Z_2$ symmetry renders the lightest $Z_2$-odd particle stable, and if it is electrically neutral then it can play the role of the DM particle. Since $m_{\chi_1}\leq m_N$, doublet fermion DM can not take place 
in this model\footnote{Furthermore, since $N$ has a direct coupling to the $Z$ gauge boson which gives rise to a spin-independent cross section orders of magnitude larger than present limits, 
it is excluded as a viable DM candidate.}. Therefore, only the neutral $Z_2$-odd scalars, either $H^0$ or $A^0$, can be the DM candidates. 
This makes this model to resemble up to some extent the IDM from the DM phenomenology point of view.
Accordingly, two possible scenarios emerge depending on whether the particles not belonging to the IDM ($S^{\pm}$, $\chi_{1,2}$ and $N$) participate or not in the DM annihilation. 
When these particles do take part of DM annihilation, the extra (not present in the IDM) coannihilation processes are the ones mediated by the Yukawa-couplings $\eta_i$, $f_i$ and $\rho_i$, 
and by the scalar couplings $\mu$ and $\lambda_6$.

For the scenario without the extra coannihilation processes, the DM phenomenology is expected to be similar to that of the IDM by assuming $m_{\kappa_2^+},\,m_{\chi_{1}}\gg m_{\kappa_1^+}$, 
a small scalar mixing angle and $\eta_i, f_i, \rho_i, \lambda_6\ll 1$. In addition, $\mu/v=\sin2\delta(m^2_{\kappa_2^+}-m^2_{\kappa_1^+})/\sqrt{2}v^2\ll1$ must also be satisfied. 
In this way, the coannihilation effects of the mentioned particles with the DM particle can be neglected. Note that the requirement of having small Yukawa-couplings is also in agreement with neutrino masses 
and $\mu\to e \gamma$ as it will be shown below. 
It follows that the viable DM mass range for this scenario (the same of the one in the IDM) is composed by
two regions \cite{Barbieri:2006dq,LopezHonorez:2006gr,Honorez:2010re,LopezHonorez:2010tb,Goudelis:2013uca,Garcia-Cely:2013zga,Arhrib:2013ela}\footnote{Without loss of generality we assume $H^0$ to be the DM candidate.}: 
the low mass regime, $m_{H^0} \simeq m_{h}/2$, and the high mass regime, $m_{H^0}\gtrsim 500$ GeV. In the region 100 GeV $\lesssim m_{H^0}<500$ GeV the gauge interactions become large so that it is not possible 
to reach the observed relic density, {\it i.e.} $\Omega_{H^0}<\Omega_{DM}$.  
In the Higgs funnel region, DM self-annihilations through the Higgs $s$-channel exchange provide the dominant contribution to the DM annihilation cross section, with $\lambda_L$ and $m_{H^0}$ as the relevant parameters.
LEP measurements give rise to the following constraints: $m_{H^0}+m_{A^0}>M_Z$, max$(m_{H^0},m_{A^0})>100$ GeV and $m_{\kappa_1^+}\gtrsim70$ GeV. 
On the other hand, for DM masses larger than 500 GeV the relic abundance strongly depends on the mass splittings between  $H^0,\,A^0$ and $\kappa_1^\pm$. 
Indeed, a small splitting of at most 15 GeV is required to reproduce the correct relic density implying that coannihilations between those particles must be taken into account.

Regarding the scenario where $S^{\pm}$, $\chi_{1,2}$ and $N$ contribute to the DM annihilation, the extra coannihilation processes involve the following initial states: 
$H_2^0\chi_{i}$, $H_2^0\kappa_i^\pm$, $\chi_i\kappa_j^\pm$, $N\kappa_j^\pm$, $\kappa_i^\pm\kappa_j^\pm$. These processes might play the main role in the calculation of the DM relic density affecting in a sensible 
way the expectations for DM detection \cite{Griest:1990kh,Profumo:2006bx,Klasen:2013jpa} and, therefore, modifying the viable parameter space of the model. 
Since a detailed analysis of the impact of these extra coannihilation channels on the relic density is beyond the scope of this work, in what follows we will no longer consider this scenario. 

\section{Constraints}
\label{sec:constraints}
\subsection{Electroweak precision tests}
\label{sec:STU}
In the present model, the new fields may modify the vacuum polarization of gauge bosons whose effects are parametrized by the $S$, $T$ and $U$ electroweak parameters~\cite{Peskin:1991sw}. 
The new fermion  ($S_F$, $T_F$) and scalar ($S_S$, $T_S$) contributions to the $S$ and $T$
parameters are~\cite{Cynolter:2008ea,Cynolter:2009my,Okada:2014qsa}\footnote{Because the $U$ parameter is suppressed by the new physics scale $U \sim \left( M_W/\Lambda \right)^2 T$,  
we do not take it into account \cite{Barbieri:2004qk}.}:
\begin{align}
  \label{eq:stu}
  S_F & = \frac{1}{3\pi} \Bigg[ 2s_\alpha^2c_\alpha^2 \left[ 1 - 3\Theta_S\left(m_{\chi_1}^2,m_{\chi_2}^2\right)  \right]
    + \log\left(\frac{m_{\chi_2}^2}{m_N^2}\right) 
    + c_\alpha^2 \log\left(\frac{m_{\chi_1}^2}{m_{\chi_2}^2}\right) \Bigg]\;, \\
  T_F & = \frac{1}{4\pi m_W^2 s_W^2} \Bigg[ \frac{\left( m_{\chi_1} - m_{\chi_2} \right)^2}{2} \left[ 
     2 c_\alpha^4 \log\left(\frac{m_{\chi_2}^2}{m_Nm_{\chi_1}}\right) + c_\alpha^2 \log\left(\frac{m_N^2}{m_{\chi_2}^2}\right) +
      c_\alpha^6 \log\left(\frac{m_{\chi_1}^2}{m_{\chi_2}^2}\right) \right] 
 \nonumber      \\
     & + 2c_\alpha^2 \Theta_T\left(m_{\chi_1}^2,m_N^2\right)  + 2s_\alpha^2 \Theta_T\left(m_{\chi_2}^2,m_N^2\right)
      - 2s_\alpha^2c_\alpha^2 \Theta_T\left(m_{\chi_1}^2,m_{\chi_2}^2\right)  \Bigg]\;,   \\
 S_S & = \frac{1}{4\pi m_Z^2}\Bigg[c_\delta^2\left(c_\delta^2-2\right)\Theta\left(m_Z^2;m_{\kappa_1^+},m_{\kappa_1^+}\right)
     + \Theta\left(m_Z^2;m_{H^0},m_{A^0}\right)  - \Theta\left(0;m_{H^0},m_{A^0}\right)  \nonumber \\
   & + s_\delta^2\left(s_\delta^2-2\right)\Theta\left(m_Z^2;m_{\kappa_2^+},m_{\kappa_2^+}\right)  
     + 2s_\delta^2c_\delta^2 
   \left[\Theta\left(m_Z^2;m_{\kappa_1^+},m_{\kappa_2^+}\right)-\Theta\left(0;m_{\kappa_1^+},m_{\kappa_2^+}\right)\right] \Bigg] \;,
    \\   
  T_S & = \frac{1}{16\pi m_W^2 s_W^2} \Bigg[ 
     c^2_\delta \Theta\left(0;m_{\kappa_1^+}, m_{H^0}\right) 
      +  c^2_\delta \Theta\left(0;m_{\kappa_1^+},m_{A^0}\right) + s^2_\delta \Theta\left(0;m_{\kappa_2^+}, m_{H^0}\right) \nonumber\\
     & +s^2_\delta \Theta\left(0;m_{\kappa_2^+}, m_{A^0}\right)
      -\Theta\left(0;m_{A^0}, m_{H^0}\right)-2s^2_\delta c^2_\delta \Theta\left(0;m_{\kappa_1^+}, m_{\kappa_2^+}\right) \Bigg]\;, \label{eq:stu2}   
\end{align}
where $c_\alpha=\cos\alpha$, $s_\alpha=\sin\alpha$, $c_\delta=\cos\delta$, $s_\delta=\sin\delta$ and the loop functions $\Theta$ are given in the Appendix \ref{sec:STUapen}. 
From these expressions we can see that the fermion contributions to $T_F$ and $S_F$ vanish in the limiting case of $\alpha=0$, which points out to the existence of a custodial symmetry.
For that reason we do not expect large deviations on $S$ and $T$ for a small mixing angle $\alpha$. 
In contrast, the scalar contributions do not tend to zero for $\delta=0$ due to the fact that after the electroweak symmetry breaking the components of the $Z_2$-odd doublet $H_2$ have mass splittings 
that are independent of $\delta$. 
However, the agreement with electroweak precision tests is reached due to the small mass splitting between $A^0$ and $\kappa_1^\pm$ ($H^0,\,A^0$ and $\kappa_1^\pm$) in the low (high) mass regime, 
just as it happens in the IDM.

\subsection{$\mu \rightarrow e \gamma$}
\label{sec:LFV}
\begin{figure}[h]
\centering
\includegraphics[scale=0.6]{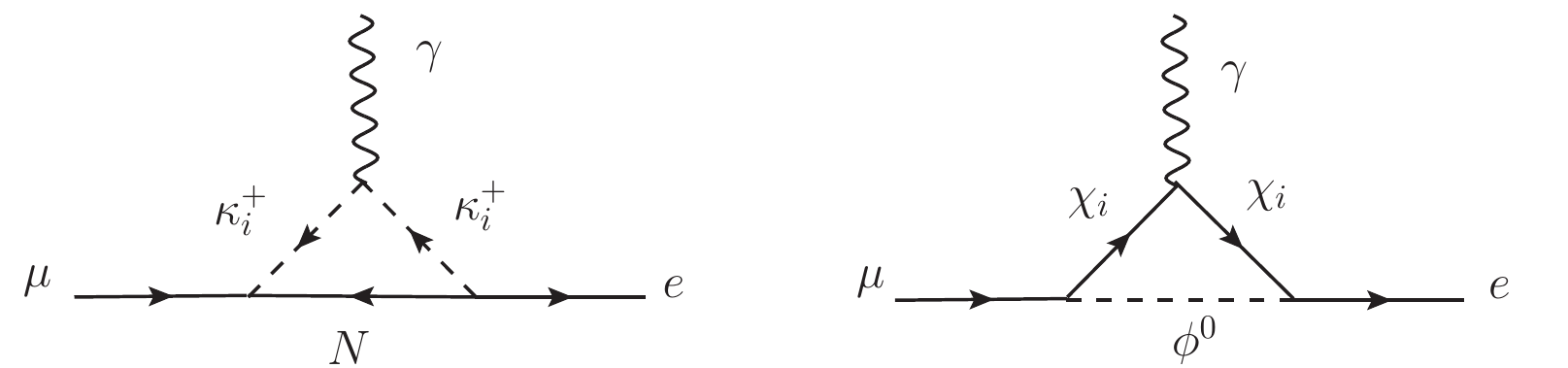}

\caption{One-loop diagrams contributing to $\mu \rightarrow e \gamma$. In the right diagram $\phi^0$ denotes the two $Z_2$-odd neutral scalars $A^0$ and $H^0$.}
\label{fig:muegammadiagrams}
\end{figure}

Lepton flavor violation processes could be a clear signal of new physics. However, due to the lack of any signal in this sector, very stringent constraints over the branching ratios for particular processes are set, 
with  $\mu \rightarrow e \gamma$ being one of the most constraining processes.
In this model such a process is controlled by the $\eta_{1,2}$, $f_{1,2}$ and $\rho_{1,2}$ Yukawa-couplings and mediated by the $Z_2$-odd particles. 
Certainly, the interactions in Eq. \eqref{eq:YukawaLagrangian} and the scalar mixing term allow to construct the one-loop diagram shown in Fig.~\ref{fig:muegammadiagrams}. 
The branching ratio for $\mu \rightarrow e \gamma$ process reads
\begin{align}
\label{eq:bramuegamma}
 \mathcal{B}\left(\mu \rightarrow e \gamma \right)
 = \frac{3\alpha_{em}}{64\pi m_{\mu}^2G_F^2} \left( \left|\Sigma_L\right|^2 + \left|\Sigma_R\right|^2 \right),
\end{align}
where $\alpha_{em}$ is the electromagnetic fine structure constant, $G_F$ is the Fermi constant and $\Sigma_L$, $\Sigma_R$ are given by
\begin{align}\label{eq:sigmaL}
 \Sigma_L =&  - \rho_1^* \eta_2^* s_\alpha c_\alpha\left[m_{\chi_1} \mathcal{G}_1(m^2_{\chi_1},m_{A^0}^2,m_{H^0}^2)-m_{\chi_2}\mathcal{G}_1(m^2_{\chi_2},m_{A^0}^2,m_{H^0}^2)\right] \nonumber\\
	    & - m_{\mu} \rho_1 \rho_2^* \left[s^2_\alpha\mathcal{F}_1(m^2_{\chi_2},m_{A^0}^2,m_{H^0}^2)+c^2_\alpha\mathcal{F}_1(m^2_{\chi_1},m_{A^0}^2,m_{H^0}^2)\right]\nonumber\\
&+m_{\mu} \rho_1 \rho_2^* \left[c^2_\delta\mathcal{F}_2(m^2_{\kappa_1^+},m_{N}^2)+s^2_\delta\mathcal{F}_2(m^2_{\kappa_2^+},m_{N}^2)\right] \;, \\
\Sigma_R =&  - \rho_2 \eta_1 s_\alpha c_\alpha\left[m_{\chi_1} \mathcal{G}_1(m^2_{\chi_1},m_{A^0}^2,m_{H^0}^2)-m_{\chi_2}\mathcal{G}_1(m^2_{\chi_2},m_{A^0}^2,m_{H^0}^2)\right] \nonumber \\
            & -m_{\mu} \eta_1 \eta_2^*\left[c^2_\alpha\mathcal{F}_1(m^2_{\chi_2},m_{A^0}^2,m_{H^0}^2) +s^2_\alpha\mathcal{F}_1(m^2_{\chi_1},m_{A^0}^2,m_{H^0}^2)\right]\nonumber\\
&+m_{\mu}f_1 f_2^*\left[s^2_\delta\mathcal{F}_2(m^2_{\kappa_1^+},m_{N}^2)+c^2_\delta\mathcal{F}_2(m^2_{\kappa_2^+},m_{N}^2)\right]\;\label{eq:sigmaR}.
\end{align}
The loop functions are presented in the Appendix \ref{sec:Loopmuegama}. 
Note that, due to the equation \eqref{eq:yukacouplings}, the couplings $\eta_2, \eta_3, f_1, f_2, f_3$ are related with $\eta_1$, 
hence, the only free Yukawa parameters entering in the expression for $\mathcal{B}(\mu \rightarrow e \gamma)$ are $\eta_1,\, \rho_1$, and $\rho_2$.

\section{Numerical results and discussion}
\label{sec:numer-results-disc}
In order to illustrate the compatibility of the model with the experimental constraints, we consider the scenario without the extra annihilation channels discussed on section \ref{sec:model}. 
Furthermore, we set $H^0$ to be the DM candidate and assume a small mixing angle $\delta$ and the mass spectrum 
with the lightest charged scalar $\kappa_1^\pm$ 
mainly doublet\footnote{It is worth mentioning that when the lightest state $\kappa_1^{\pm}$ is mainly singlet, the relic density cannot be obtained without considering the coannihilation processes
with $\kappa_2^{\pm}$ unless that $m_{\kappa_1^\pm}\gtrsim 300$ GeV and $m_{H^0}\simeq m_h/2$, in which case the relic density is independent of $m_{\kappa_{2}^{\pm}}$.}.

For the low mass regime and without lose of generality we assume $m_{\kappa_1^+},m_{A^0}>100$ GeV and $|\delta|\lesssim0.2$, which implies that the remaining $Z_2$-odd fields do not alter the DM phenomenology 
expected for the IDM in that regime. On the other hand, to quantitatively assess up to what extent the presence of the new fermion fields and $\kappa_2^\pm$ could affect the expected phenomenology in the
high mass regime, through the opening of new (co-)annihilation channels, we have calculated the DM relic density through micrOMEGAs \cite{Belanger:2013oya} via FeynRules \cite{Alloul:2013bka} and
make a scan (to be described below) over the free parameters of the model. For this purpose, we have set $\lambda_2,\,\lambda_S$ and all the Yukawa-couplings to $10^{-2}$. 
The numerical result confirms the preliminary expectations: when $m_{\kappa_2^+}/m_{\kappa_1^+}\gtrsim 1.1$, $|\delta|\lesssim 0.2$ and $|\mu|/v\lesssim 10^{-1}$ 
the new (co-)annihilations channels compared with those present in the IDM do not play a significant role in the determination of DM relic density. 

\begin{figure}[htb]
\centering
\includegraphics[scale=0.45]{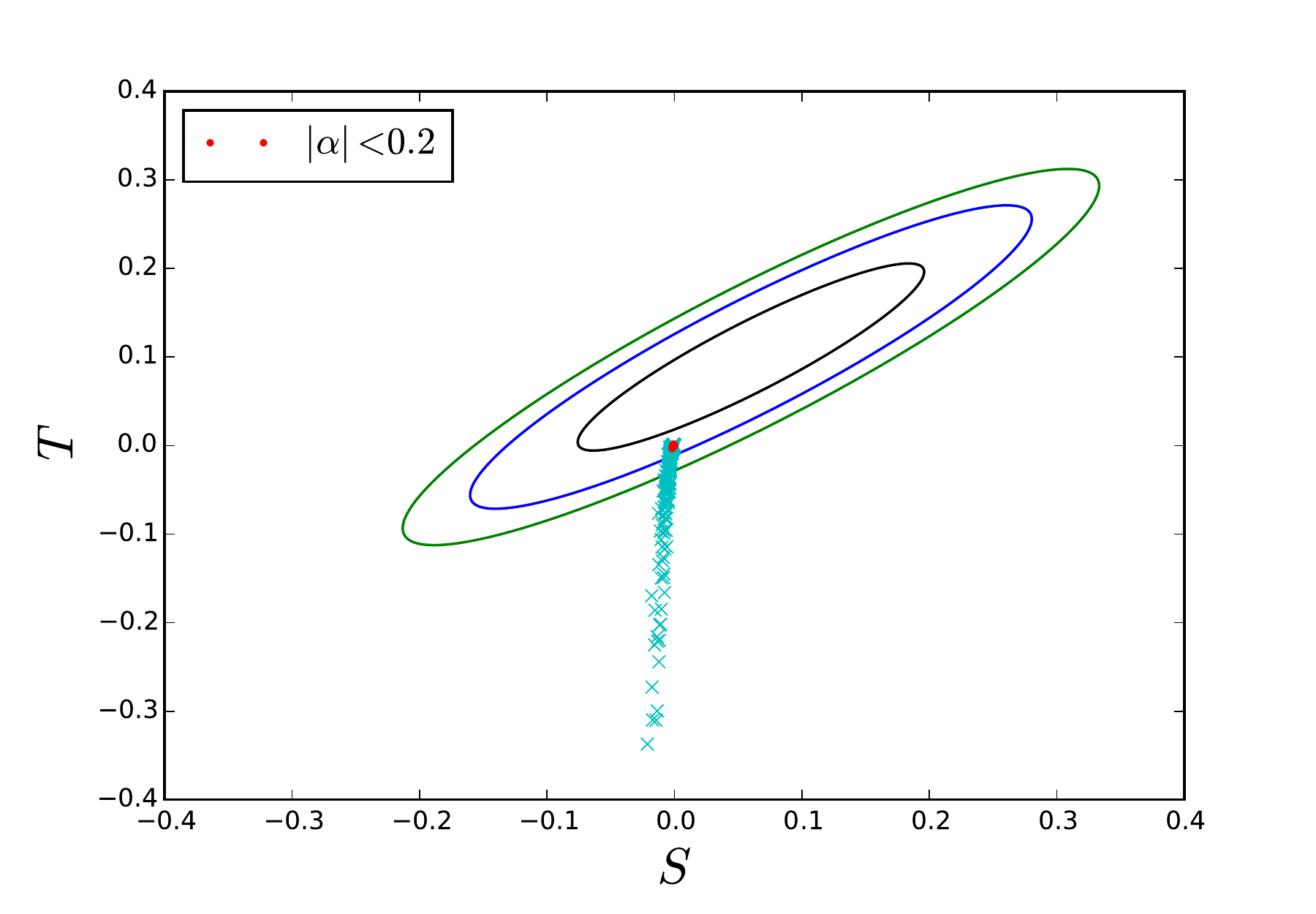}
\includegraphics[scale=0.45]{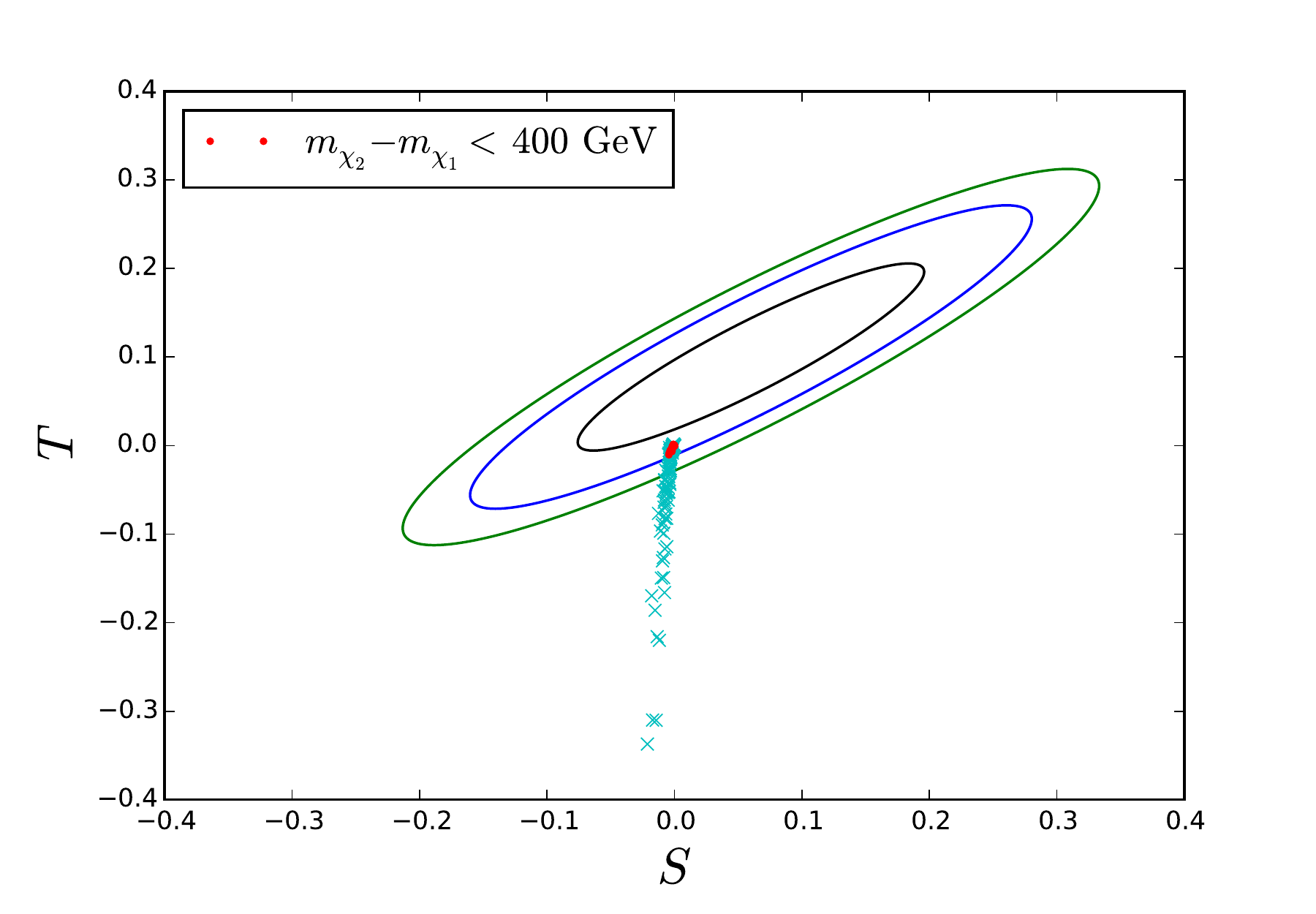}
\caption{Contour plot for $Z_2$-odd scalar and fermion contributions to the EWPT parameters in the high mass regime. 
The left panel shows the $S$, $T$ contributions for any mass splitting $m_{\chi_2} - m_{\chi_1}$,
while the right panel shows 
$S$, $T$ contributions for any value of the mixing angle $\alpha$.}
\label{fig:highmassregime-ST}
\end{figure}

Regarding the electroweak precision test, we have performed a numerical analysis for the two DM mass regimes mentioned above. For the high mass regime, we have considered  the following ranges for the free parameters:
\begin{align}
\label{eq:scan}
500\, {\rm GeV} < m_{H^0}< 1\, {\rm TeV}\;;\; 
m_{A^0},\, m_{\kappa_1^+} = m_{H^0} + [0.1,10]\, {\rm GeV}\;; \nonumber \\
m_{\kappa_2^+} = m_{\kappa_1^+}  + [0.1,1000]\, {\rm GeV}\;;\;
m_{\chi_1} = m_{\kappa_2^+}  + [0.1,1000]\, {\rm GeV}\;; \nonumber \\
m_{\chi_2} = m_{\chi_1}  + [0.1,1000]\, {\rm GeV}\;; \;\delta,\alpha \in [-\pi/2,\pi/2]\;;\;\lambda_L,\lambda_6 \in [10^{-3},1].
\end{align}
The scalar and fermion contributions to $S$ and $T$ are shown in Fig.~\ref{fig:highmassregime-ST}, where the constraints coming from the DM phenomenology mentioned above have been taken into account. 
The black, blue and green ellipses represent the experimental constraints at 68\% CL, 95\% CL and 99\% CL, 
respectively \cite{Baak:2014ora}\footnote{The experimental deviations from the SM predictions in the $S$ and $T$ parameters for 
$m_h = 126$ GeV, $m_t = 173 $ GeV and $U=0$ are $S = 0.06 \pm 0.09\;,\; T = 0.10 \pm 0.07$ where the correlation factor between 
$S$ and $T$ is $0.91$ \cite{Baak:2014ora}.}. 
It is worth to mention 
that contrary to the IDM, in our model the $S$ and $T$ parameters are not negligible in the high mass regime because the 
fermion contributions are already present. However, the constraints are easily satisfied for a small fermion mixing angle
$|\alpha|\lesssim0.2$ (red points in the left-panel). 
On the other hand, by allowing arbitrary values for the mixing angle, $\alpha$, the contributions to $S$ and $T$ 
are kept within the $2\sigma$ level as long as $m_{\chi_2} - m_{\chi_1} \lesssim 400$~GeV (red points in the
right-panel). 

\begin{figure}[b]
\centering
\includegraphics[scale=0.5]{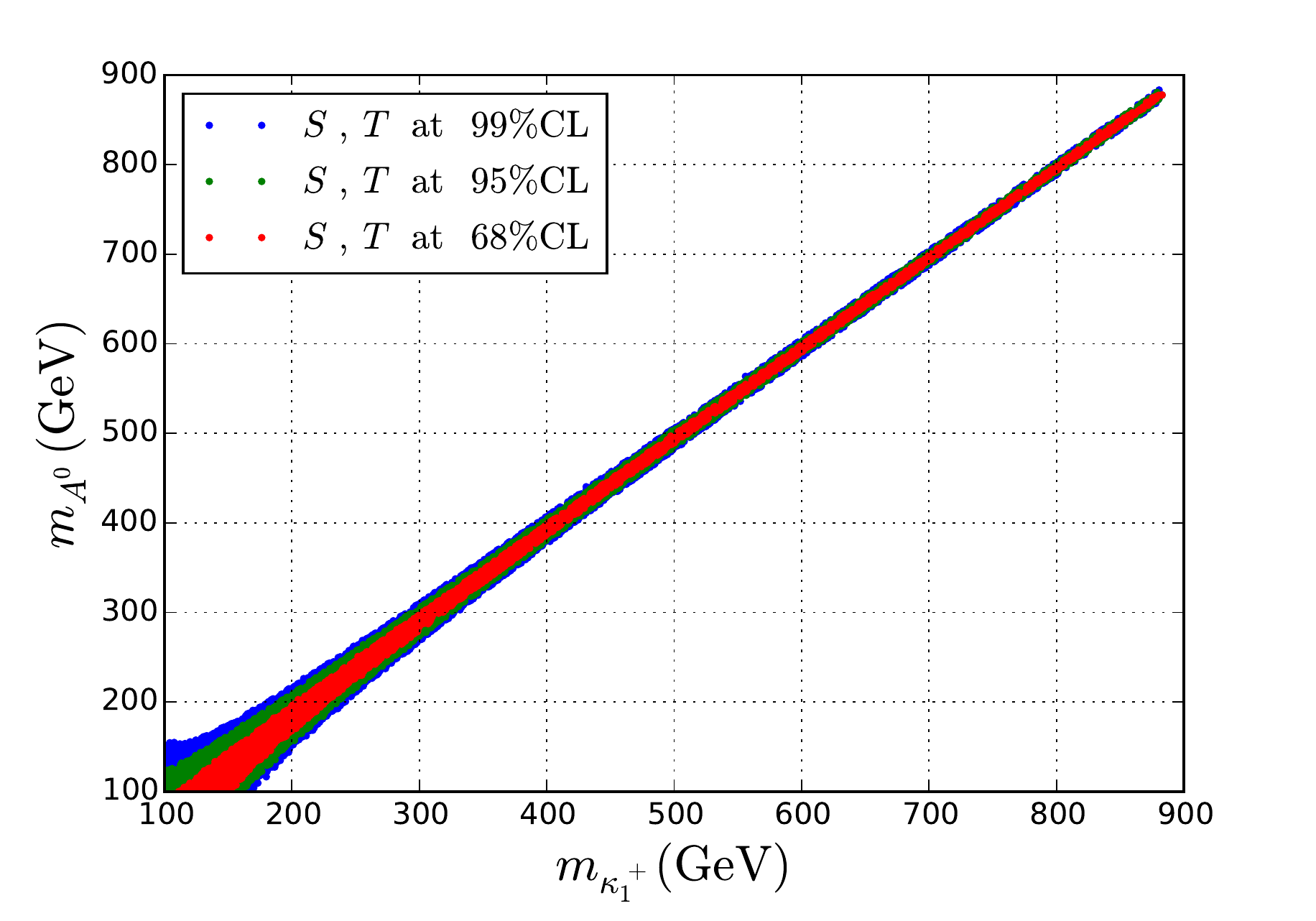}
\caption{$S$, $T$ constraints on the masses $m_{A^0}$ and $m_{\kappa_1^+}$. We have taken $|\alpha|<0.1$ and $|\delta|<0.1$.}
\label{fig:mA-mk1-STUlow}
\end{figure}

Regarding the low mass regime we have varied the free parameters as follows: 
$60\, {\rm GeV} < m_{H^0}< 80\, {\rm GeV}$, $100\, {\rm GeV}< m_{A^0},m_{\kappa_1^+} < 1000\, {\rm GeV}$, 
$m_{\kappa_1^+}<m_{\kappa_2^+}<m_{\chi_1} <1000\, {\rm GeV}$, and the same ranges in the Eq. \eqref{eq:scan} for 
the mixing angles and scalar couplings. 
The fermion contributions to $S$ and $T$ are satisfied by imposing either $|\alpha|\lesssim 0.1$ 
or $m_{\chi_2} - m_{\chi_1} \lesssim 200$~GeV.
In this case, the scalar contributions are not kept within the $2\sigma$ level by just imposing the DM phenomenology of the IDM.
This occurs because in the low mass regime there is always a non-negligible mass splitting between the 
DM particle and the other scalars. 
Fig.~\ref{fig:mA-mk1-STUlow} shows the allowed values for the masses 
$m_{A^0}$ and $m_{\kappa_1^+}$ that satisfy the $S$, $T$ parameters at 68\% CL (red points), 
95\% CL (green points) and 99\% CL (blue points) respectively. 
We have taken $|\alpha|\lesssim0.1$ in order to suppress the fermion contribution.
Note that if $m_{A^0}$ is increased, $m_{\kappa_1^+}$ will have to be increased. However, from the 
unitary constraints given in Eq. \eqref{eq:contraints_scalarparameters} 
an upper limit is obtained on the scalar masses, which leads to that they should be nearly degenerate at 800 GeV.

\begin{figure}[htb]
\centering
\includegraphics[scale=0.5]{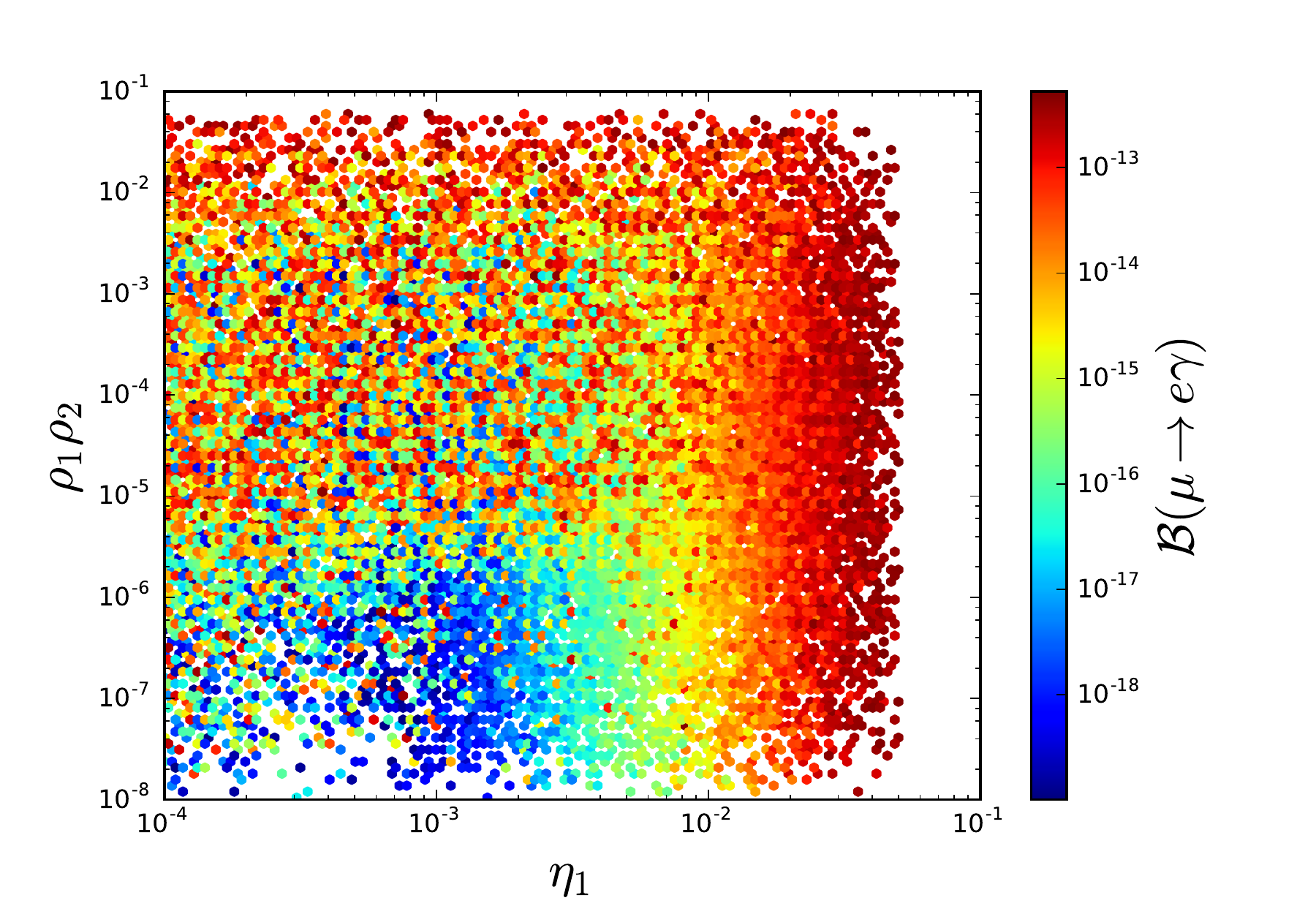}
\caption{
Region in the ($\eta_1$, $\rho_1 \rho_2$) plane for the high mass regime which is compatible with the current $\mu\rightarrow e \gamma$ constraint. 
Note that the correlation between $\rho_1\rho_2$ and $\eta$ can be spoiled by the interference between the two $\rho_1\rho_2$ contributions and/or
by the $\rho_1^*\eta_2^*$ and $\rho_2\eta_1$ contributions (see Eqs. (25) and (26)), since it is possible to obtain low values of $\mathcal{B}(\mu\rightarrow e\gamma)$ (color code) 
with relative large values of $\rho_1\rho_2\sim 10^{-2}$. 
For a inverted hierarchy and $\lambda =-1$ the numerical results are similiar to those for the normal hierarchy: $\rho_1\rho_2\leq0.08$ and $\eta_1\leq0.3$ for the current bound. 
}
\label{fig:LFVyukawas}
\end{figure}

Concerning to the LFV constraints, we have focused on the current strongest bound,
which is provided by $\mu\rightarrow e \gamma$ process.
We have made a scan over the free parameters of the model for the CP-conserving scenario (the CP Dirac phase is fixed to zero) with a normal hierarchy and choosing 
$\lambda=-1$.
For this purpose, we have varied the free parameters within the ranges given in Eq. \eqref{eq:scan}, 
in addition to $\eta_1,\rho_1, \rho_2,\in [10^{-4},1]$.
The results are shown in Fig.~\ref{fig:LFVyukawas}. 
All the points satisfy the current bound \cite{Adam:2013mnn} and only a minority will be probed by future searches~\cite{Baldini:2013ke}.
We have taken $|\alpha|\lesssim 0.1$, $m_{\kappa_2^+}/m_{\kappa_1^+}\gtrsim 1.1$, $|\delta|\lesssim 0.2$ 
and $|\mu|/v\lesssim 10^{-1}$ in order to satisfy the oblique parameters and preserve the DM phenomenology expected for the IDM.
Note that the $\mathcal{B}(\mu \rightarrow e\gamma)$ limit can be easily satisfied imposing 
$\rho_1 \rho_2  \lesssim 4\times 10^{-2} $ and $\eta_1 \lesssim 5\times 10^{-2} $. 
On the other hand, for the low mass regime we obtain similar results to those in the high mass regime. 
Remember that, in order to satisfy the oblique parameter we need to impose small mixing angles as well as a nearly degenerate masses 
between $A^0$ and $\kappa_{1}^+$.

Finally, we turn  the discussion to collider searches. The high-mass region of the IDM
is quite difficult to probe at the LHC. However, the low mass region can be probed by searching for dilepton plus missing transverse energy
signal \cite{Dolle:2009ft,Belanger:2015kga,Ilnicka:2015jba} and trilepton plus missing transverse energy signal \cite{Miao:2010rg,Abe:2014gua} 
with a sensitivity in the parameter region with $\kappa_1^+, A^0$ 100–180~GeV. A similar sensitivity could be expected for $\kappa_2^+$.

Concerning VL fermions, the searches performed at LEPII
impose a limit of $m_{\chi_1} > $ 100 GeV~\cite{Achard:2001qw}.
At the LHC, the larger exclusion for VL fermion is expected for large mass splittings, 100\% branching ratios to electron or muons, 
and higher  fermions $SU(2)_L$ representations. 
In our case, it  corresponds to a higgsino-like VL fermion production without final state taus.  
For example, if a higgsino-like charged fermion is the next to lightest $Z_2$-odd particle and choosing the Yukawa-couplings such that
\begin{align}
\operatorname{max}\left(\rho_1,\rho_2\right)\gg \operatorname{min}\left(\rho_1,\rho_2\right),\rho_3,\eta_i\,,
\end{align}
we have a dilepton plus missing transverse energy signal from
\begin{align}
pp\to& \chi_1^+\chi_1^- \to l^+ l^-  H^0 H^0\,, & l^\pm=&e^\pm,\mu^\pm\,.
\end{align}
Since the cuts for this kind of signal  at  the LHC (in both ATLAS and CMS) do not depend in angular distributions between the final states, 
the corresponding excluded cross sections  are insensitive to the spin of the produced particles.  Currently, they are  interpreted in terms of  slepton pair production.
A recast of the excluded cross section for slepton pair production $pp\rightarrow \tilde{l}^+\tilde{l}^-\rightarrow l^+l^-\tilde{\chi}_1^0\tilde{\chi}_1^0$, 
studied in Ref.~\cite{Aad:2014vma}\footnote{Where the lightest neutralino, $\tilde{\chi}_1^0$, is the dark matter candidate.},  
allows to exclude higgsino-like charged fermions up to $510$~GeV~\cite{Restrepo:2015ura}.

Conversely, in the case of $\chi_1^\pm$ nearly degenerate with $H^0$ (compressed spectra), the bounds on $m_{\chi_1}$ are $\sim 100\ \text{GeV}$
for $\Delta m=m_{\chi_1}-m_{H^0}<50\ \text{GeV}$~\cite{Khachatryan:2015kxa}\footnote{In our case, 
the low mass region of the IDM  with $m_{H^0}=70\ $GeV combined with the LEPII constraint on VL fermions, imply that $\Delta m>30\ \text{GeV}$.}.
If, in addition, the Yukawa-couplings are such that 
\begin{align}
\rho_3\gg \rho_1,\rho_2,\eta_i\,,
\end{align}
then $\mathcal{B}\left( \chi^\pm \to \tau H^0 \right)\approx1$, and the exclusion limits are worse due to the larger $\tau$ misidentification rates.  
Recently, an extended analysis of  the LHC Run-I data have been presented by ATLAS~\cite{Aad:2015eda} with new searches for compressed spectra and final state taus. 
In particular, by using multivariate analysis techniques, the 95\% excluded cross section for
 $pp\rightarrow \tilde{\tau}_{R,L}^+\tilde{\tau}_{R,L}^-\rightarrow \tau^+\tau^-\tilde{\chi}_1^0\tilde{\chi}_1^0$ is given for several neutralino masses. 
As expected, and in contrast to the selectron and smuon pair production, 
there is no sensitivity to left-  or right-stau pair production.
By using the same strategy than in~\cite{Preparation:2015}, we focus in the excluded cross section plot presented in Fig.~12 of Ref.~\cite{Aad:2015eda} for a DM particle of 60~GeV, 
since it is a  representative value  in the case of the  IDM to account for the  proper relic density. 
Because of the larger cross section for pair produced higgsinos decaying into two taus plus missing transverse energy, 
we are able to exclude higgsino-like charged fermions in the range $115<m_{\chi_1^+}/\text{GeV}<180$ by using the theoretical cross section calculated to next-to-leading order in~\cite{Restrepo:2015ura}.

Another attempt to circumvent both problems have been made recently  in Ref.~\cite{Khachatryan:2015kxa} of the CMS collaboration, 
by implementing the vector boson fusion topology to pair produce electroweakinos~\cite{Dutta:2012xe}.  
There, supersymmetric models with bino-like $\widetilde{\chi}_1^0$ and wino-like $\widetilde{\chi}_2^0$ and $\widetilde{\chi}_1^\pm$ are considered in the presence of a light stau. 
Assuming $\mathcal{B}\left( \widetilde{\chi}_1^\pm\to \nu\widetilde{\tau}^{\pm}\to \nu\tau^\pm\widetilde{\chi}_1^0 \right)=1$ and $\mathcal{B}\left( \widetilde{\chi}_2^0\to \tau^\pm\widetilde{\tau}^{\mp}\to \tau^\pm\tau^\mp\widetilde{\chi}_1^0 \right)=1$,
they are able to find some supersymmetric scenarios where the LEPII constraint can be improved.  
We could expect that a similar analysis for the higgsino-like charged VL fermion may allow to close the previous gap until around  $115\ \text{GeV}$. 
A detailed recast of this CMS analysis, will be done elsewhere. In summary, we expect an exclusion for the higgsino-like charged VL fermions of the model around 180~GeV.
On the other hand, searches in the di-tau plus missing transverse energy signature have been studied in Ref.~\cite{Yu:2014mfa}. There, it was 
shown that the high luminosity LHC of 3000 fb$^{-1}$  can exclude $SU(2)_L$-singlet charged VL fermion  up to  $m_{\chi_1}\sim 450$~GeV. 

\section{Conclusions}
\label{sec:conclusions}

We have considered an extension of the Zee model which involves two vector-like leptons, 
a doublet and a singlet of $SU(2)_L$ and the imposition of an exact $Z_2$ symmetry. This symmetry, under which all the non-Standard Model fields are odd, avoids tree-level Higgs-mediated flavor
changing neutral currents and ensures the stability of the lightest neutral component inside the second scalar doublet and, therefore, allowing to have a viable dark matter candidate.
We have shown that under some conditions the well-known DM phenomenology of the IDM is recovered. 
As in the Zee model, neutrino masses are generated at one loop,
leading to either a normal mass hierarchy or a inverted mass hierarchy. 
However, due to the flavor structure of the neutrino mass matrix, one neutrino remains massless. 
Moreover, such a flavor structure always allows to reproduce the correct neutrino oscillation parameters and to have only four free Yukawa-couplings (of a total of nine), 
which can be constrained using the $\mu \rightarrow e \gamma$ lepton flavor violation process. 
In particular, we have found that $\rho_1 \rho_2  \lesssim 10^{-2}$ and $\eta_1 \lesssim 10^{-2} $ in order to fulfill that constraint.
On the other hand, the oblique parameters impose $|\alpha|\lesssim 0.2$ and $m_{\chi_2} - m_{\chi_1} \lesssim 400$ GeV for the high mass regime 
while $|\alpha|\lesssim 0.1$ and $m_{\chi_2} - m_{\chi_1} \lesssim 200$ GeV for the low mass regime.
Finally, we argued that in general, the collider limits for vector-like leptons are not so far from the limit imposed by LEPII.

\section*{Acknowledgments}

We are very gratefully to Federico von der Pahlen for illuminating
discussions.
D. R. and O. Z. are supported by UdeA through the grants Sostenibilidad-GFIF, CODI-2014-361 and CODI-IN650CE, and COLCIENCIAS
through the grants numbers 111-556-934918 and 111-565-842691. D. P. and R. L. are supported by COLCIENCIAS. R. L. acknowledges the hospitality of Universidade Federal do ABC in the final stage of this work.

\appendix{}

\section{Free parameters}
\label{sec:freeparameters}
Some of the scalar potential parameters can be written in terms of physical scalar masses using the relations in \eqref{eq:scalar_neutralsmasss} and \eqref{eq:schmasses}:
\begin{align}
 \frac{1}{2}v^2\lambda_3 & = 
 m_{\kappa_1^+}^2\cos^2\delta + m_{\kappa_2^+}^2\sin^2\delta - m_{H^0}^2 + v^2\lambda_L \;,\qquad
 \mu_S^2 =  m_{\kappa_1^+}^2\sin^2\delta + m_{\kappa_2^+}^2\cos^2\delta - \frac{1}{2}v^2\lambda_6 \;,\nonumber \\ 
  v^2\lambda_5 &= m_{H^0}^2-m_{A^0}^2\;, \qquad
  \lambda_4  = 2\lambda_L - \lambda_3 - \lambda_5\;,\qquad \mu_2^2 = m_{H^0}^2 - \lambda_Lv^2 \;.
\end{align}
\section{ST formulae}
\label{sec:STUapen}
Here, we present the analytical loop functions used for the analysis of the $S$ and $T$ parameters,
\begin{align}
 \Theta_S(m_1,m_2)& = \frac{2}{9} + \frac{(m_1^2 + m_2^2)(m_1^4 - 4m_1^2m_2^2 + m_2^4) + 6m_1^3m_2^3}{6(m_1^2 - m_2^2)^3}
		\log\left(\frac{m_1^2}{m_2^2}\right) 
	    + \frac{4m_1^2m_2^2 - 3m_1m_2(m_1^2+m_2^2)}{6(m_1^2 - m_2^2)^2} \;, \nonumber \\
 \Theta_T(m_1,m_2)& = \frac{m_1^2+m_2^2}{4} - \frac{(m_1^4 - 2m_1m_2(m_1^2+m_2^2) + m_2^4)}{4(m_1^2 - m_2^2)}
		\log\left(\frac{m_2^2}{m_1^2}\right),
 \nonumber \\
 \Theta(p^2;m_1,m_2)&=\int_0^1 dx \Big[(2x-1)(m_1^2-m_2^2)+(2x-1)^2p^2\Big]\ln \Big[xm_1^2+(1-x)m_2^2-x(1-x)p^2\Big],\nonumber \\
 \Theta(0;m_1,m_2)&=\frac{m_1^2+m_2^2}{2}-\frac{m_1^2m_2^2}{m_1^2-m_2^2}\ln\left( \frac{m_1^2}{m_2^2} \right).
\end{align}

\section{Loop function in the $\mu\to e\gamma$}
\label{sec:Loopmuegama}
Here, we present the analytical loop functions used for the analysis of the $\mu \rightarrow e\gamma$ constraint,
\begin{align}
\mathcal{G}_1(m_a^2,m_b^2,m_c^2)&=\frac{1}{m_b^2} G\left(\frac{m_a^2}{m_b^2}\right)-\frac{1}{m_c^2} G\left(\frac{m_a^2}{m_c^2}\right),\\
\mathcal{F}_1(m^2_{a},m_{b}^2,m_{c}^2)&=\frac{1}{2 m_{a}^2} \left[ F\left(\frac{m_{b}^2}{m_{a}^2}\right)+ F\left(\frac{m_{c}^2}{m_{a}^2}\right) \right],\\
\mathcal{F}_2(m^2_{a},m_{b}^2)&=\mathcal{F}_1(m_{a}^2,m_{b}^2,m_{b}^2),
\end{align}
where
\begin{align}
 F\left(x\right) = \frac{2x^3 + 3x^2 -6x +1 -6x^2\log\left(x\right)}{6\left(x-1\right)^4}\,,\qquad 
 G\left(x\right) = \frac{x^2 -4x + 3 + 2\log\left(x\right)}{2\left(x-1\right)^3}\,.
\end{align}
\bibliographystyle{h-physrev4}
\bibliography{darkmatter}
\end{document}